# Web Mining using Artificial Ant Colonies : A Survey

Richa Gupta
*Department of Computer Science*
*University of Delhi*

**ABSTRACT :** *Web mining has been very crucial to any organization as it provides useful insights to business patterns. It helps the company to understand its customers better. As the web is growing in pace, so is its importance and hence it becomes all the more necessary to find useful patterns. Here in this paper, web mining using ant colony optimization has been reviewed with some of its experimental results.*

***Keyword-*** *web mining, artificial ant colonies*

## 1. INTRODUCTION

Web mining is the application of data mining techniques to discover usage patterns from large Web data repositories. It is the extraction of interesting and useful knowledge and implicit information from artifacts or activity related to the WWW. It may reveal interesting and unknown knowledge about both users and websites which could be used by different systems for analysis. It is used to understand customer behaviour, evaluate the effectiveness of a particular Web site, and help quantify the success of a marketing campaign [1,5].

## 2. IMPORTANCE

In world where everything is moving on a fast pace, new business starting each day and they deploying the internet marketing potential, one needs to know more and more about the marketing needs. It becomes important to study the behaviour of how web sites are visited. It provides with business intelligence data that can be helpful in determining the behaviour of customer and his needs. It helps in establishing relation between various variable of market analysis. Web mining involves taking as input large data and provide certain insights into useful business behaviours [6,7].

Web data mining also helps in predictive analysis. These are used to analyse current and historical data for making predictions about the future sales of potential growth [6].

Web mining provides with many benefits in many fields. In the field of Financial analysis, it includes reviewing the costs and revenues, calculations and comparative analysis of corporate income statements, comparative analysis of corporate income statements, cash flow statement, analysis of financial markets etc. In the field of Marketing analysis it includes analysis of sales receipts, sales profitability, profit margins, sales targets, stock exchange quotations and market identification and segmentation. In the field of Customer analysis, it includes customer profitability, modelling customer behaviour and reactions, customer satisfaction etc. Web mining in this field helps us to find strategy that should be used to get number of customers with quality [3].

Web mining has certain generic goals:

- Understanding customer's behaviour and experience.
- Determining effective logical structure for the website
- Improving business supported by the website
- Building user profiles by combining user's navigation paths with other features like page viewing time, page content, hyperlink structure

## 3. WEB MINING TASKS

The steps for mining the web are:

1. Resource location – it includes finding the required Web documents, information and services. The data can be collected from three main sources – web servers, proxy servers and web clients [3].
2. Information extraction – it means extracting the desired information from data source
3. Generalization – it includes finding general patterns at individual websites as well as across multiple websites.
4. Analysis – it includes validation and verification of the discovered patterns.

## 4. AREAS OF WEB MINING

There are three different areas of Web mining based on the type of data used

1. Web Content Mining
2. Web Structure Mining
3. Web Usage Mining

**Web Content Mining**

It is the process of extracting useful information and knowledge from the Web contents/data/documents. Content may consist of text, images, audio, video or structured records such as lists and tables. Most of the Web content mining methods are based on unstructured free text data or structured HTML documents. Specific area of Web content mining is *text data mining*. Web content mining is differentiated from two different points of view: Information Retrieval View and Database View [8].

**Web Structure Mining**

It is the process of using graph theory to analyse the node and connection structure of a website. The model reflects the topology of the hyper-links underlying the website. It can be used to generate information on the *similarity or the difference between different websites*. According to web structural data, web structure mining can be divided into two kinds:

- Extracting patterns from hyperlinks in the web
- Mining the document structure, that is, analysis of the tree-like structure of web pages [5].

**Web Usage Mining or Log Mining**

It is the process of extracting useful knowledge from the data obtained from Web user sessions. Some users might be looking at only textual data, whereas some others might be interested in multimedia data. It tries to find usage patterns from the Web data to understand and better serve the needs of Web-based applications. Web sessions capture the identity or origin of web users along with their browsing behaviour at a website. Some applications of Web usage mining are *adaptive websites, web personalization and recommendation, business intelligence* etc. Web usage mining can itself be classified further depending on the kind of usage data:

- Web Server Data – it has the user logs collected by the web server. Typically data includes IP address, page reference and access time.
- Application Server Data – it tracks various kinds of business events and log them in application server logs.
- Application Level Data – new kinds of events can be defined in an application and logging can be turned on for them thus generating histories of these specially defined events [5].

A generic Web usage mining framework is shown in fig below

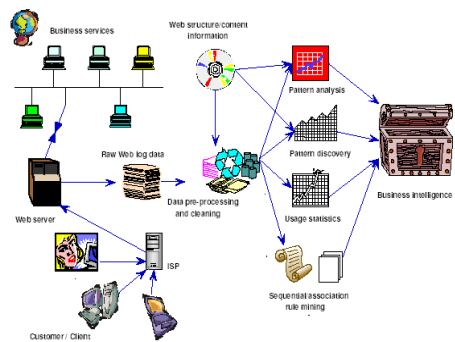

Figure 1: Web usage mining framework

## 5. WEB SESSIONS

The Web usage mining mainly relies on the study of Web Sessions. These sessions are extracted from Web server log files. They reflect the activity of a user on a Web site during a given period of time. They are very noisy, since the user is generally not clearly identified. When the user identification is unknown, the requests contained in the log files are sorted by IP number and by date, and then grouped according to the estimated time that lasts a session. Several methods have been developed to extract knowledge from the sessions to describe the navigational behaviours of the users on a Website.

The underlying idea of the method is that the algorithm should gather in the same cluster the sessions that corresponds to the users that navigate similarly i.e that have the same motivation and interests.

Each session extracted from the Web server log file contains the following parameters:

1) The IP number of the user
2) The identity of the user (if known)
3) The date and time of the connection
4) The reconstructed history that lists sequentially the web pages that have been visited during the session
5) The representation of session as a transaction vector which states for each of page of the web site if it has been accessed at least one time during the session
6) A time vector that contains for each page, the estimated time that user spent. For the last page seen, the time is estimated by the mean of the times spent on the other pages, as it cannot be computed. For the pages that are loaded too quickly, it is possible to have a time equal to 0. in this case, we artificially set this time to 1 second to be consistent with the number of impacts recorded for each page
7) A date vector that lists for each page of the web site, the first date they have been visited
8) A hits vector, that counts for each page of the web site, the number of hits that have been recorded
9) The total time of the session

## 6. ANT COLONY OPTIMIZATION

**Defining ACO**

Ant Colony Optimization (ACO) is a technique for optimization that was introduced in nearly 1990s. The inspiration of ACO algorithms are the behaviour of real ant colonies.

The complex social behaviours of ants have been much studied by Science, and computer scientists are now finding that these behaviour patterns can provide models for solving difficult combinatorial optimization problems. They attempt to develop algorithms inspired by various aspects of ant behaviour [9].

**Real Ant Behaviours**

The social behaviour of ants is among the most complex in the insect world. They communicate by touching and smelling their odour. There are many behaviours of ants through which they communicate and perform different tasks. Some of those behaviours are described below:

**Ant Foraging**

Ants form and maintain a line to their food source by laying a trail of pheromone, i.e. a chemical to which other members of the same species are very sensitive. They deposit a certain amount of pheromone while walking, and each ant prefers to follow a direction rich in pheromone. This enables the ant colony to quickly find the shortest route. The first ants to return should normally be those on the shortest route, so this will be the first to be doubly marked by pheromone (once in each direction). Thus other ants will be more attracted to this route than to longer ones not yet doubly marked, which means it will become even more strongly marked with pheromone. Soon, therefore, nearly all the ants will choose this route.

These algorithms form the heart of a new research field called "***Ant Optimization***". They are applied to problems like the ***Traveling Salesman*** and related ***network-based problems***.

**Division of Labour and Task Allocation**

An ant colony is able to adjust its task allocation to external perturbation. It maintains its functions, even when a huge partition of (specialized) workers is artificially removed. This adaptation apparently occurs without any central control, conscious evaluation of the global situation or direct communication. A model that reproduces this colony level behaviour based on self-organization deals with response thresholds. Every ant perceives the need for the fulfilment of some task - brood care or storage of incoming food. If this stimulus is strong enough - it is increasing when the task is not carried out - the ant has a higher probability to engage in the task associated with the stimulus. Ants with low thresholds for one task respond on lower levels of stimuli. Having terminated a task successfully, the ant may adapt its threshold for that task. This reinforcement of the bias to repeat kinds of work that the ant has often done before, leads to specialization in a population of formally identical workers.

This model of "foraging for work" can serve as an basis for a ***task allocation mechanism in an***

*artificial multi-agent system*. Possible applications for online differentiation, e.g. as a *distributed control algorithm for robots* are sketched.

**Cementry Organization and Brood Sorting**

Another self-organized behaviour observed in ant colonies is the aggregation of corpses or larval sorting. Distributed objects are loaded and deposited in clusters without any central plan indicating where to unload the items. A simple agent-based model explains this phenomenon. An agent carrying an item drops it when it perceives many similar objects. Conversely, an agent without any load picks up an item when it perceives that most objects in its surroundings are of a different kind.

This model can be directly implemented in a swarm of robots driving around and moving pucks. It also can provide a mechanism for *data analysis or graph partitioning*.

**Colonial Odor**

Every day, real ants solve a crucial recognition problem when they meet. They have to decide to which nest they belong in order to guarantee the survival of the nest. This phenomenon is known as "**colonial closure**". It relies on continuous exchange and updation of chemical cues on their cuticle. Each ant has own view of its colony odor at given time, and updates it continuously. Ant in this way preserves its nest from being attacked by predators or parasite and reinforces its integration of the nest.

These artificial ants are able to construct group of similar objects, a problem which is known as *data clustering*.

## 7. CLUSTERING WEB SESSIONS USING ANT COLONY OPTIMIZATION

The basic idea is to apply clustering algorithm to web sessions to discover homogeneous groups of sessions, that is, to gather in same cluster the sessions that correspond to the users that navigate similarly, that is to say, that have the same motivation and interests.

The colonial closure behaviour of ants (already described) can be applied to clustering the web sessions. Each web session is considered as one ant. Each ant has to distinguish its nestmates from intruders. They achieve this goal by detecting odor that is spread over the cuticle of the encountered ants (called *label*) and then compares this perceived *label* to a neuronal *template*.

ANTCLUST is one the algorithm which can be used for clustering of web session logs [7].

**AntClust Parameters**

Artificial ants can be considered as a set of parameters that evolve according to behavioural rules [7].

As presented in [7], For each ant *i*, the parameters are defined as follows

- *Label$_i$* - indicates the belonging nest of the ant and is simply coded by a number. At the beginning of the algorithm, the ant does not belong to a nest, so *Label$_i$* = 0. The label evolves until the ant finds the nest that best corresponds to its genome.
- *Genome$_i$* - corresponds to the object of the data set. It is not modified during the algorithm. When the ants meet, they compare their *genome* to evaluate their similarity. For eg: we can use web session as *genome*.
- *Template$_i$* or *T$_i$* - it is an acceptance threshold that is coded by a real value between 0 and 1. It is learned during an initialization period in which artificial ant *i* meets other ants, and each time evaluates the similarity between their *genomes*.

$T_i \Leftarrow [ Sim(i, .) + Max(Sim(i, .)) ] / 2$

Once artificial ants have learned their template, they use it during their meetings to decide if they should accept the encountered ants. The *acceptance* mechanism between two ants *i* and *j* is defined as follows:

$A(i,j) \Leftrightarrow (Sim(i,j) > T_i)$ and $(Sim(i,j) > T_j)$

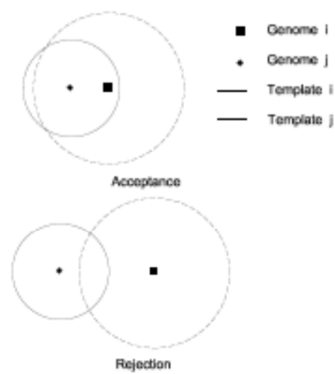

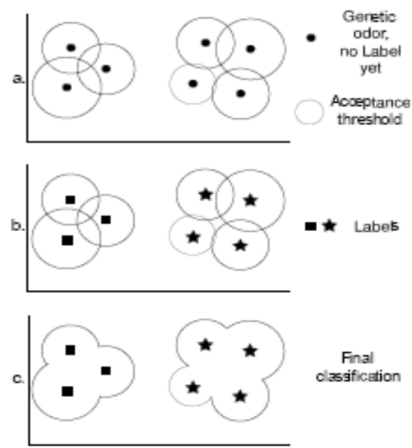

Figure 1: Principle of acceptance and rejection between ants $i$ and $j$

### Behavioural Rules

R1 New nest creation:

If ($Label_i = Label_j = 0$) and $Acceptance(i,j)$ then create a new Label $Label_{NEW}$ and $Label_i \leftarrow Label_{NEW}$, $Label_j \leftarrow Label_{NEW}$. If Acceptance is false then nothing happens.

R2 Adding an ant with no Label to an existing ant:

If ($Label_i = 0$ and $Label_j \neq 0$) and $Acceptance(i,j)$. Then $Label_i \leftarrow Label_j$. The case ($Label_j = 0$ and $Label_i \neq 0$) is handled in a similar way.

R3 Meeting between two nestmates:

If ($Label_i \neq Label_j$) and $Acceptance(i,j)$. Then the ant belonging to the smaller nest changes its nest and belongs now to the nest of the encountered ant.

### AntClust Algorithm

1) Initialization of the ants:
2) for all ants i $\in$ [1, n]
3) Genomei $\leftarrow$ ith session of the data set
4) Labeli $\leftarrow$ 0
5) Templatei is learned
6) NbITER $\leftarrow$ 75 * n
7) Simuate NbITER iterations during which two randomly chosen ants meet
8) Assign each ant having no more nest to the nest of the most similar ant found that have a nest

### Experimental Results

The algorithm is evaluated over the server logs of Delhi University. The performance and running time of algorithm is recorded and analyzed over different sets of test data. The algorithm is able to handle large data sets with an affordable computing time.

### Complexity

Assuming the following parameters:

$n$: Number of transactions in the Web log file.

$N$: Number of Web Sessions generated.

The complexity of Pre-processing step is O($n$)

The complexity of Initialization Phase is O($n^2$) and that of Simulation Phase is O($N^2$). Hence the complexity of the code is O($n^2+N^2$). Since $n>>N$, the worst case complexity can be given as O($n^2$).

### Experiments and Results

The code was run on multiple data sets and its performance has been recorded as follows:

From the above table the average number of sessions and clusters for each run can be given as follows:

| Transactions in Log file | Average number of Sessions | Average number of Clusters |
|---|---|---|
| 5,000 | 129 | 10 |
| 10,000 | 231 | 11 |
| 20,000 | 396 | 10 |
| 30,000 | 582 | 11 |
| 40,000 | 761 | 13 |
| 50,000 | 917 | 12 |

The figure, shows the graph between transactions in log file and sessions.

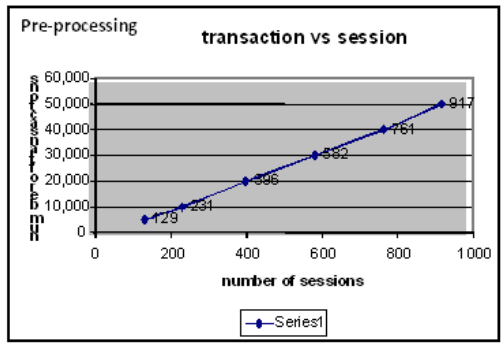

Figure 2: Transaction-Session graph

From the figure above, it can be seen that transactions in log file are directly proportional to the number of sessions generated. It is true for a popular website which is accessed by different users.

The next graph is a plot between number of sessions and clusters generated.

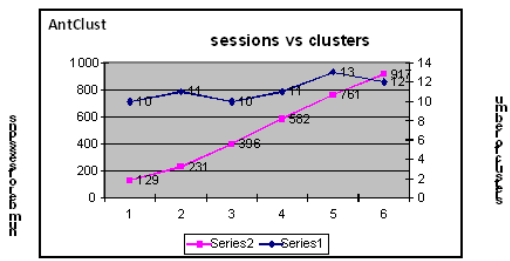

Figure 3: Sessions vs clusters

From Figure 3, it can be observed that the increase in number of sessions do not directly correspond to the increase in number of clusters.

### 8. CONCLUSION

Web mining is a promising research topic which provides useful business solutions. We stated the types of web mining and the importance of it in detail. Also, Ant Colony Optimization has been explained and how it can be used for Web Usage Mining.

When applied on Web Sessions, AntClust finds 189 clusters. The 4 dominating clusters contain more than half of the sessions. The partitions generated can prove useful to analyze the navigation pattern of users and understanding the interests of the Web users. AntClust spends approximately 19.05 minutes at 3.00GHz, to cluster the User Sessions, which is affordable time.